\RequirePackage{fix-cm}
\documentclass[twocolumn]{svjour3}          % twocolumn
\smartqed  % flush right qed marks, e.g. at end of proof
\usepackage{lipsum}
\usepackage{mathtools}
\usepackage{cuted}
\usepackage{graphicx}
\usepackage{bbding}
\usepackage{upgreek}
\usepackage{url}
\usepackage{color,xcolor}
\usepackage{amsmath}
\usepackage{bm}
\usepackage[colorlinks,
            linkcolor=blue,
            anchorcolor=blue,
            citecolor=blue,
            urlcolor=blue,
            ]{hyperref}
\usepackage{titletoc}
\usepackage{amssymb}
\usepackage{flushend}
\usepackage[pagewise]{lineno}

\usepackage{latexsym}

\usepackage{epsf}
\usepackage{float}

\usepackage{epsfig}
\usepackage{subfigure}
\usepackage[flushleft]{threeparttable}

\usepackage{tikz}
\usetikzlibrary{shapes}
\definecolor{blue}{RGB}{0,112,192}
\definecolor{lightblue}{RGB}{0,176,240}
\definecolor{green}{RGB}{0,176,80}
\definecolor{yellow}{RGB}{255,255,0}
\definecolor{orange}{RGB}{255,192,0}
\definecolor{red}{RGB}{255,0,0}
\definecolor{darkred}{RGB}{118,0,0}
\definecolor{purple}{RGB}{208,0,154}
\emergencystretch=\maxdimen
\titlecontents{section}[0pt]{\addvspace{2pt}\filright}              {\contentspush{\thecontentslabel\ }}              {}{\titlerule*[8pt]{.}\contentspage}

\hypersetup{
    bookmarks=false,         % show bookmarks bar?
    colorlinks=true,       % false: boxed links; true: colored links
    linkcolor=blue,          % color of internal links (change box color with linkbordercolor)
    linkbordercolor={1 1 1}
}

\journalname{Acta Mech. Sin.}
\begin{document}
%\begin{sloppypar}
\rmfamily
\title{Friction-dependent rheology of dry granular systems%\thanks{Grants or other notes
}
%\subtitle{({\it Acta Mechanica Sinica})}

\titlerunning{Friction-dependent rheology of dry granular systems}        % if too long for running head

\author{Teng Man$^{1*}$ \and Pei Zhang$^{1*}$ \and Zhuan Ge$^{1,2}$ \and Sergio A. Galindo-Torres$^1$ \and K. M. Hill$^3$%\and
       }


\institute{{\Envelope} Sergio A. Galindo-Torres \at
            \email{s.torres@westlake.edu.cn} \at \at
            %{\Envelope} T. Man \at
            % \email{manteng@westlake.edu.cn} \at \at
            $^{*}$ T. Man and P. Zhang have equal contribution to this work.\at
		    $^{1}$ Key Laboratory of Coastal Environment and Resources of Zhejiang Province (KLaCER), School of Engineering, Westlake University, 18 Shilongshan St., Hangzhou, Zhejiang 310024, China \at
		    $^{2}$ College of Civil Engineering and Architecture, Zhejiang University, 866 Yuhangtang Rd., Hangzhou, Zhejiang 310058, China \at
            $^{3}$ Saint Anthony Fall Laboratory and the Department of Civil, Environmental, and Geo- Engineering, University of Minnesota, 500 Pillsbury Dr. S.E., Minneapolis, MN 55455, United States
           }
%Received: date / Accepted: date / Published online: 7 February 2020\\
\date{\copyright {\it Acta Mechanica Sinica}, The Chinese Society of Theoretical and Applied Mechanics (CSTAM) 2020
}

%\contentsline{section}{\numberline{12.8}Some section that is wrapped in the TOC}{87}

\maketitle

%\tableofcontents
%\linenumbers
%\modulolinenumbers[5]
\begin{abstract}
Understanding the rheology of granular assemblies is important for natural and engineering systems, but the relationship between inter-particle friction (or microscopic friction) and macroscopic friction is still not well understood. In this study, using the the discrete element method (DEM) with spherical particles and realistic contact laws, we investigate the mechanics of granular systems with a wide range of inter-particle frictional coefficients and aim to establish a friction-dependent rheology for dry granular flows. The corresponding results show that increasing inter-particle friction dramatically increases the effective frictional coefficient, $\mu_{\textrm{eff}}$, while decreasing the solid fraction of the system and increasing the transitional inertial number that marks the division of quasi-static regimes and intermediate flow regimes. We further propose a new dimensionless number, $\mathcal{M}$, as a ratio between the inertial effect and frictional effect, which is similar to the effective aspect ratio in granular column collapses, and unifies the influence of inter-particle friction with the inertial number. We then establish a relationship between $\mathcal{M}$ and the dimensionless granular temperature, $\Theta$, to further universalize the influence of inter-particle frictions. Such study can broaden the application of the $\mu(I)$ rheology in natural and engineering systems and help establish a more general constitutive model for complex granular systems.
\end{abstract}
\vspace{1 cm}

\section{\label{sec:intro}Introduction}

Granular systems are frequently encountered and commonly used in both nature and engineering systems. Understanding the rheological behavior of them is crucial to civil engineering, chemical engineering, pharmaceutical engineering, food processing, and geophysical flows \cite{guyon2020built}. The attempt to accurately describe stresses of granular flows has a long history and can date back to Bagnold's work \cite{bagnold1954}, where he stated that, for granular systems in inertial regimes, both their shear stress and pressure can be written as $f_{\tau, p}(\phi_s)\rho_p\dot{\gamma}^2 d^2$, where $f_{\tau}(\cdot)$ and $f_p(\cdot)$ stand for functions of the shear stress and the pressure, $\phi_s$ is the solid fraction of the system, $\rho_p$ is the average particle density, $\dot{\gamma}$ is the shear rate, and $d$ is the average particle diameter. Lun and Savage \cite{lun1984kinetic,lun1991kinetic}, and Jenkins \cite{jenkins2002kinetic} et al developed a kinetic theory for granular systems which can capture well the behavior of granular gases and can be extended to dense systems\cite{berzi2020extended}.

Based on the Bagnold rheology \cite{Silbert2001gran}, we witness tremendous advances in understanding the basic governing principles, especially the constitutive relationships, of granular flows in past decades\cite{POULIQUEN2002163,pouliquen2002}. Among these investigations, the proposal of the $\mu(I)$ rheology \cite{jop2006} and the $\mu(I, I_v)$ relationship \cite{trulsson2012,man2022two} of granular materials opened a window for exploring the behavior of dense granular materials from a viewpoint of the competition among acting stresses, viscous stresses and inertial stresses \cite{trulsson2012,forterre2008flows} (or the competition among different time scales $1/\dot{\gamma}$, $d/\sqrt{\sigma_n/\rho_p}$, and $\eta_f/\sigma_n$ \cite{cassar2005}, where $\sigma_n$ is the pressure and $\eta_f$ the dynamic viscosity of the interstitial fluid).

The current $\mu(I)$ rheology states that the effective frictional coefficient, $\mu_{\textrm{eff}} = \tau/\sigma_n$, where $\tau$ is the shear stress, depends on the inertial number, $I = \dot{\gamma}d/\sqrt{\sigma_n/\rho_p}$ with the relationship \cite{pouliquen2002,midi2004,pouliquen2006}
\begin{equation} \label{eq-muI}
\begin{split}
    \mu_{\textrm{eff}} = \mu_s + \frac{\mu_1 - \mu_s}{1 + I_0/I},
\end{split}
\end{equation}
where $\mu_s$, $\mu_1$, and $I_0$ are fitting parameter, but $\mu_s$ and $\mu_1$ can be seen as critical frictional coefficients at zero shear rate and high inertial number, respectively, and $I_0$ can be seen as the transitional inertial number that divided the system into quasi-static regime and intermediate flow regime. Kamrin et al. \cite{Kamrin2012nonlocal} and Bouzid et al. \cite{Bouzid2013} later considered the nonlocal effect when describing the constitutive relationship of granular flows, and Kim and Kamrin \cite{Kim2020power} proposed a power-law scaling for steady-state granular systems with various boundary condition, with $\mu_{\textrm{eff}}$ scaling with both $I$ and $\Theta = \rho_{p}T_{g}/\sigma_n$, where $\Theta$ is a dimensionless granular temperature, $T_g = \delta v^2/D$ is the granular temperature, $D$ is the space dimension, and $\delta v^2$ is the squared velocity fluctuation of the system.

Some research concerned about the influence of inter-particle friction on the macroscopic behavior of granular systems \cite{Peyneau2008solid,Hatano2007,Azema2015internal}, and found that, (i) frictionless granular flows of spheres (or disks, polygons, etc.) still had an effective frictional coefficient $\mu_{\textrm{eff}} \approx 0.1$ when $I\rightarrow 0$ \cite{Peyneau2008solid,Azema2018inertial}, (ii) increasing inter-particle friction, $\mu_p$, led to an increase in $\mu_{\textrm{eff}}$ but the relationship between $\mu_{\textrm{eff}}$ and $\mu_p$ was nonlinear \cite{Hatano2007,Hatano2010}, and (iii) once $\mu_p>0.3$, the influence of inter-particle friction became insignificant. Srivastava et al \cite{srivastava2022flow,clemmer2021shear} also reported the influence of inter-particle friction on the constitutive behavior of dry granular systems in their extensive investigation of the flow, arrest, and rate-dependent effects of stressed granular systems. Based on $I$ and $\mu_p$, DeGiuli et al. \cite{degiuli2016} classified inertial granular flows into three different regimes: frictionless, frictional sliding, and rolling; they further analyzed the differences in kinetics and rheology for systems in different regimes. However, no constitutive relationship is obtained to quantitatively incorporate the influence of inter-particle frictions, which limits the application of $\mu(I)$ rheology in real natural and engineering systems.

In this study, we use the discrete element method (DEM) to study the simple shear simulation of granular systems with a wide range of inter-particle frictional coefficients to investigate the influence of microscopic friction on macroscopic friction. The interaction between particles is represented as Hertzian-Mindlin contact law with energy dissipation. We first investigated the influence of inter-particle friction on the classical $\mu-I$ rheology and proposed a new dimensionless number to represent the influence of interparticle frictions. Additionally, We introduce the dimensionless granular temperature to consider fluctuations introduced by the inter-particle friction so that we could establish a constitutive relationship for steady-state granular systems. In the end, we discussed the influence of interparticle friction on the transition of granular assemblies from a metastable state to a liquid-like state, as well as the influence of interparticle friction on the contact statistics in a sheared granular system.

\section{Methodology}

\subsection{Contact laws}

We use DEM of an in-house code to investigate the behavior of granular materials so that we could obtain detailed particle-scale information of the system. In our simulations, both the normal and tangential contact forces are calculated based on Hertz-Mindlin contact theories \cite{cundall1979} and damping components based on the derivation outlined by Tsuji et al. \cite{tsuji1992}. In this model, the tangential contact force also follows the Coulomb friction law, where the tangential contact forces cannot exceed $\mu_p F_{c}$. Thus, the interaction between particle $i$ and particle $j$ can be written as
\begin{linenomath*}
\begin{subequations} \label{eq:contact}
\begin{align}
F_{c}^{\rm{ij,n}} = -k_n\delta_{n}^{1.5} - \eta_n\delta_{n}^{0.25}\dot{\delta}_{n}, \label{eq:contact1}\\
F_{c}^{\rm{ij,t}} = \rm{min}\big( -k_{t}\delta_{n}^{0.5}\delta_t - \eta_{t}\delta_n^{0.25}\dot{\delta}_t, \mu_{p}\it{F}_{c}^{\rm{ij,n}} \big), \label{eq:contact2}
\end{align}
\end{subequations}
\end{linenomath*}
where $F_{c}^{\rm{ij,n}}$ and $F_{c}^{\rm{ij,t}}$ are normal and tangential contact forces acting on particle \textit{i} from particle \textit{j}. $\delta_{n}$ is the overlap between particles in normal direction in DEM simulation, which is given by $\delta_{n} = R_{i}+R_{j} - |\vec{r}_{i} - \vec{r}_{j}|$, where $R_i$ and $R_j$ are particle radii, and $\vec{r}_{i}$ and $\vec{r}_{j}$ are position vectors of two particles. $\delta_t$ is the corresponding tangential deformation at the contact point between two particles, and $\mu_{p}$ is the frictional coefficient. The parameters in the contact model are related to material properties of two contacting particles presented in Table \ref{table:1}. In this simulation, the particle density is 2650 kg/m$^3$, elastic modulus is 29 GPa, and the Poisson's ratio is 0.20. The material properties are similar to those of limestone particles. In order to calculate the dissipative term, the coefficient of restitution is set to be around 0.20. Usually, in studies related to granular materials, the particles are relatively smooth and elastic. However, in this study, we chose a relatively small number for the coefficient of restitution to capture a more representative collisional behavior of grains whose collisions are less elastic \cite{foerster1994measurements,lorenz1997measurements}. The coefficient of friction between particles, $\mu_p$, ranges from $1\times 10^{-5}$ to 0.4.

\begin{table}[h]
\begin{threeparttable}
\caption{Relationships for calculating the stiffnesses and damping coefficients in Eq. \ref{eq:contact}(a) and (b)}% title of Table
\centering % used for centering table
\begin{tabular}{c c}
\hline\hline
Variables & Equations \\
\hline
$k_n$ & $(4/3)\sqrt{R_{\textrm{eff}}}E_{\textrm{eff}}$ \\
$k_t$ & $8\sqrt{R_{\textrm{eff}}}G_{\textrm{eff}}$ \\
$\eta_{n}$ & $\alpha_o \sqrt{m_{\textrm{eff}}k_{n}}$ \\
$\eta_{t}$ & $\alpha_o \sqrt{m_{\textrm{eff}}k_{t}}$ \\
$R_{\textrm{eff}}$ & $(1/R_i + 1/R_j)^{-1}$ \\
$E_{\textrm{eff}}$ & \ \ \ \ \ \ \ \ \ $[(1-\nu_{i}^2)/E_i + (1-\nu_{j}^2)/E_j]^{-1}$\ \ \ \ \ \ \ \ \  \\
$G_{\textrm{eff}}$ & $(1/G_i + 1/G_j)^{-1}$ \\
$m_{\textrm{eff}}$ & $(1/m_i + 1/m_j)^{-1}$ \\
\hline\hline
\end{tabular}
\begin{tablenotes}
      %\small
      \item $\alpha_o = 0.9$ is calculated based on the relationship between $\alpha_o$ and $e$ proposed by Tsuji et al. \cite{tsuji1992}. $E_i$ and $E_j$ are elastic moduli,  $G_i$ and $G_j$ are shear moduli, $\nu_i$ and $\nu_j$ are Poisson's ratios, and $m_i$ and $m_j$ are masses of contacting particles $i$ and $j$.
\end{tablenotes}
\label{table:1}
\end{threeparttable}
\end{table}

\subsection{Simulation setup}

\begin{figure}[!ht]
  \includegraphics[scale = 0.35]{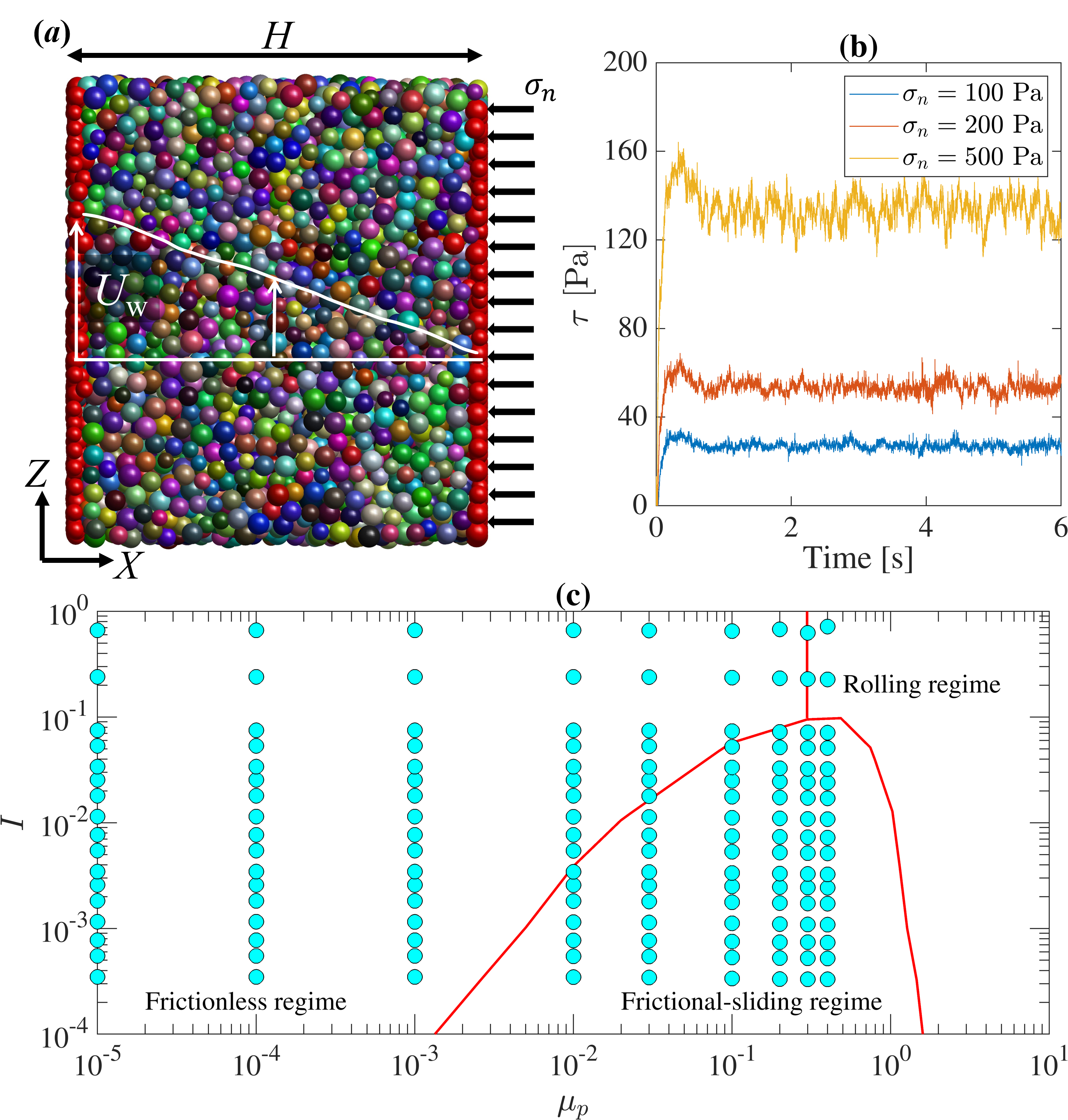}
  \caption{Simulation setup: (a) DEM simulation of simple shear tests with constant pressure and periodic boundary conditions in $Y-$ and $Z-$ directions; (b) time-evolution of shear stress of the simulation with $\dot{\gamma} \approx$ 0.56 s$^{-1}$; (c) our simulation setup plotted with the phase diagram suggested by DeGiuli et al. \cite{degiuli2016}.}
  \label{fig1}
\end{figure}

To measure the constitutive relationship of the granular system, we shear it in a rectangular box [Fig. \ref{fig1}(a)] with periodic boundary conditions in both $X-$ and $Y-$ directions. The length of the system in $X-$direction is 58.3 mm and the width of the system in $Y-$direction is 30 mm. The diameter of the particles is uniformly distributed from 2.0 mm to 2.5 mm. The system is sheared in $Z-$direction by one layer of particles glued on the boundary walls with constant boundary velocity, $U_w$, but the height of the system, $H$, varies with the constant pressure, $\sigma_n$, and boundary velocity, $U_w$. The glued particles can ensure that the system is sheared by a rough surface. Such a simple shear test with constant pressure and wall velocity leads to a linear velocity profile across the $x-$direction [Fig. \ref{fig1}(a)], which results in a relatively constant shear rate of $\dot{\gamma} = U_w/H(t)$. We measure the pressure, $\sigma_n$, the shear stress, $\tau$, and the shear rate, $\dot{\gamma}$, when the system reaches the steady-state, which is defined when the measured $\tau$ becomes stable, as shown in Fig. \ref{fig1}(b). We also calculate the inertial number, $I$, of each simulation when the system is already in steady state.

DeGiuli et al. \cite{degiuli2016} suggested that the inertial number and the inter-particle frictional coefficient together define a phase diagram, which classified granular systems into three different regimes: (i) frictionless regime, (ii) frictional-sliding regime, and (iii) rolling regime. We plot our data with the phase diagram in Fig. \ref{fig1}(c) and find that most of our work falls in the frictionless regime and the frictional-sliding regime, while only four simulations can be classified as in the rolling regime. Even though our simulations cannot fully describe the behavior of granular systems in a rolling regime, our choice of both $I$ and $\mu_p$ has a relatively wide range.

Usually, we calculate the effective frictional coefficient, $\mu_{\textrm{eff}} = \tau/\sigma_n$, and solid fraction, $\phi_s$, and relate them to the inertial number $I$. Equation \ref{eq-muI} shows the classical relationship between $\mu_{\textrm{eff}}$ and $I$, and the classical relationship between $\phi_s$ and $I$ can be written as $\phi_s = {\phi_m}/(1 +\beta_{\phi} I^{\alpha})$, where $\phi_m$ is fitted but represents the maximum solid fraction that a sheared granular system can reach, and $\beta_{\phi}$ and $\alpha$ are fitting parameters, and $\alpha$ is often equal to $1$ \cite{midi2004}. In a constant pressure condition, the shear stress comes from two sources: one source is the pure friction between contacting particles; the other source is the inter-lock effect due to normal collision between particles. Thus, changing the inter-particle friction may dramatically influence the $\mu_{\textrm{eff}}\sim I$ rheology.

\section{Results and discussions}

\begin{figure*}
  \centering
  \includegraphics[scale = 0.5]{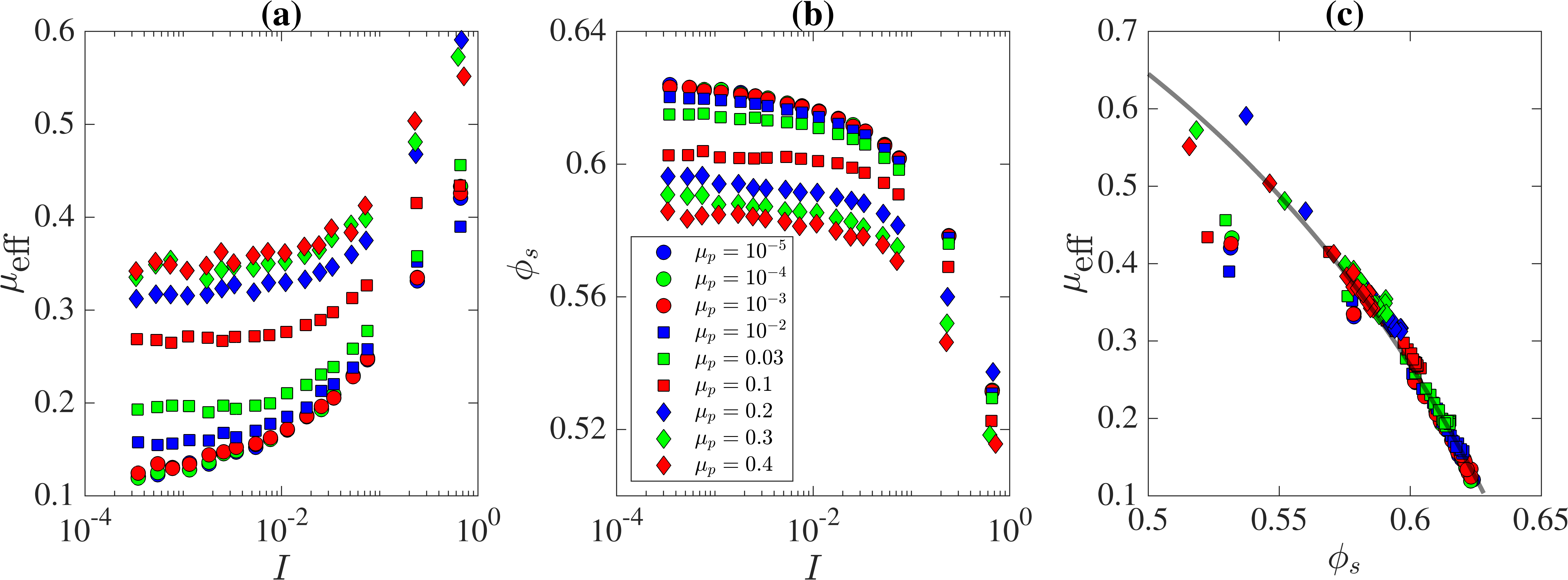}
  \caption{Simulation results: (a) The relationship between the effective aspect ratio, $\mu_{\textrm{eff}}$, and the inertial number, $I$; (b) the relationship between the solid fraction, $\phi_s$, and $I$; (c) the relationship between $\mu_{\textrm{eff}}$ and $\phi_s$}
  \label{fig2}
\end{figure*}

\subsection{Rheological behavior}

We plot both the $\mu_{\textrm{eff}}(I)$ and the $\phi_s(I)$ relationships in Fig. \ref{fig2}. Each type of markers represents simulations with the same inter-particle frictional coefficient, $\mu_p$, but different loading conditions. Considering only the frictional rheology [Fig. \ref{fig2}(a)], for simulations with the same $\mu_p$, as we vary $\sigma_n$ and $\dot{\gamma}$ to achieve different $I$'s, the $\mu_{\textrm{eff}}(I)$ relationship collapses onto one curve, and approximately follows the Eq.\ref{eq-muI}. The $\mu_{\textrm{eff}}(I)$ relationship changes as we increase the inter-particle frictional coefficient. When $\mu_p \leq 10^{-3}$, $\mu_{\textrm{eff}}$ increase from 0.12 to approximately 0.38 as we increase $I$ approximately from $8\times 10^{-4}$ to $0.66$. However, as we further increase $\mu_p$, the $\mu_{\textrm{eff}}$ shift upward accordingly. When $\mu_p = 0.4$, $\mu_{\textrm{eff}}\rightarrow 0.36$ as $I\rightarrow 0$, and $\mu_{\textrm{eff}}$ reach $\approx 0.6$ when $I$ is approaching 1. This implies that each $\mu_p$ corresponds to a distinct $\mu_s$ in Eq. \ref{eq-muI}. 

Interestingly, changing $\mu_p$ can not only influence the $\mu_{\textrm{eff}}$ but also influence the transitional inertial number. When $\mu_p \leq 10^{-3}$, $\mu_{\textrm{eff}}$ starts to increase as soon as $I$ is larger than $\approx 10^{-3}$. However, for simulations with $\mu_p = 0.4$, the $\mu_{\textrm{eff}}(I)$ curve keeps being flat until $I\approx 0.05$. We have stated in Sec. \ref{sec:intro} that the transitional inertial number marks the transition from a quasi-static flow to an intermediate flow. This suggests that changing inter-particle friction influences this regime transition, and it is reasonable since increasing inter-particle friction will increase the energy dissipation during particle collisions thus increase the shearing resistance, which can further increase the threshold for a granular system to develop into an intermediate flow.

We observe similar behavior in the $\phi_s(I)$ relationship in Fig. \ref{fig2}(b). For systems with the same $\mu_p$, the $\phi_s(I)$ relationship also collapses onto one curve. However, increase $\mu_p$ from $10^{-5}$ to 0.4 shifts the $\phi_s(I)$ relationship downward that each $\mu_p$ corresponding to a different $\phi_m$. Similar to the $\mu_{\textrm{eff}}(I)$ relationship, systems with small $\mu_p$ transition from quasi-static flow to intermediate flow at smaller inertial number than systems with large $\mu_p$.

Even though we find great influence of friction on both the $\mu_{\textrm{eff}}(I)$ and $\phi_s(I)$ relationships, The relationship between $\mu_{\textrm{eff}}$ and $\phi_s$, shown in Fig. \ref{fig2}(c), remain relatively universal. Excluding systems with $\mu_p \leq 0.1$ and $\phi_s<0.55$, other simulation results of $\mu_{\textrm{eff}}$ and $\phi_s$ collapse onto one curve. The universal $\mu_{\textrm{eff}}(\phi_s)$ relationship, which was validated by multiple previous works, indicates the reliability of our simulation results. However, we cannot neglect the deviation from this universal $\mu_{\textrm{eff}}(\phi_s)$ relationship when $\mu_p \leq 0.1$ and $\phi_s<0.55$, and have to perform detailed analysis to study the influence of friction on the rheological behavior of granular systems, and find out how changing inter-particle frictional coefficient can dramatically shift the behavior of both $\mu_{\textrm{eff}}(I)$ and $\phi_s(I)$ relationships.

\subsection{Microscopic friction and macroscopic friction}

\begin{figure}[ht]
  \includegraphics[scale = 0.35]{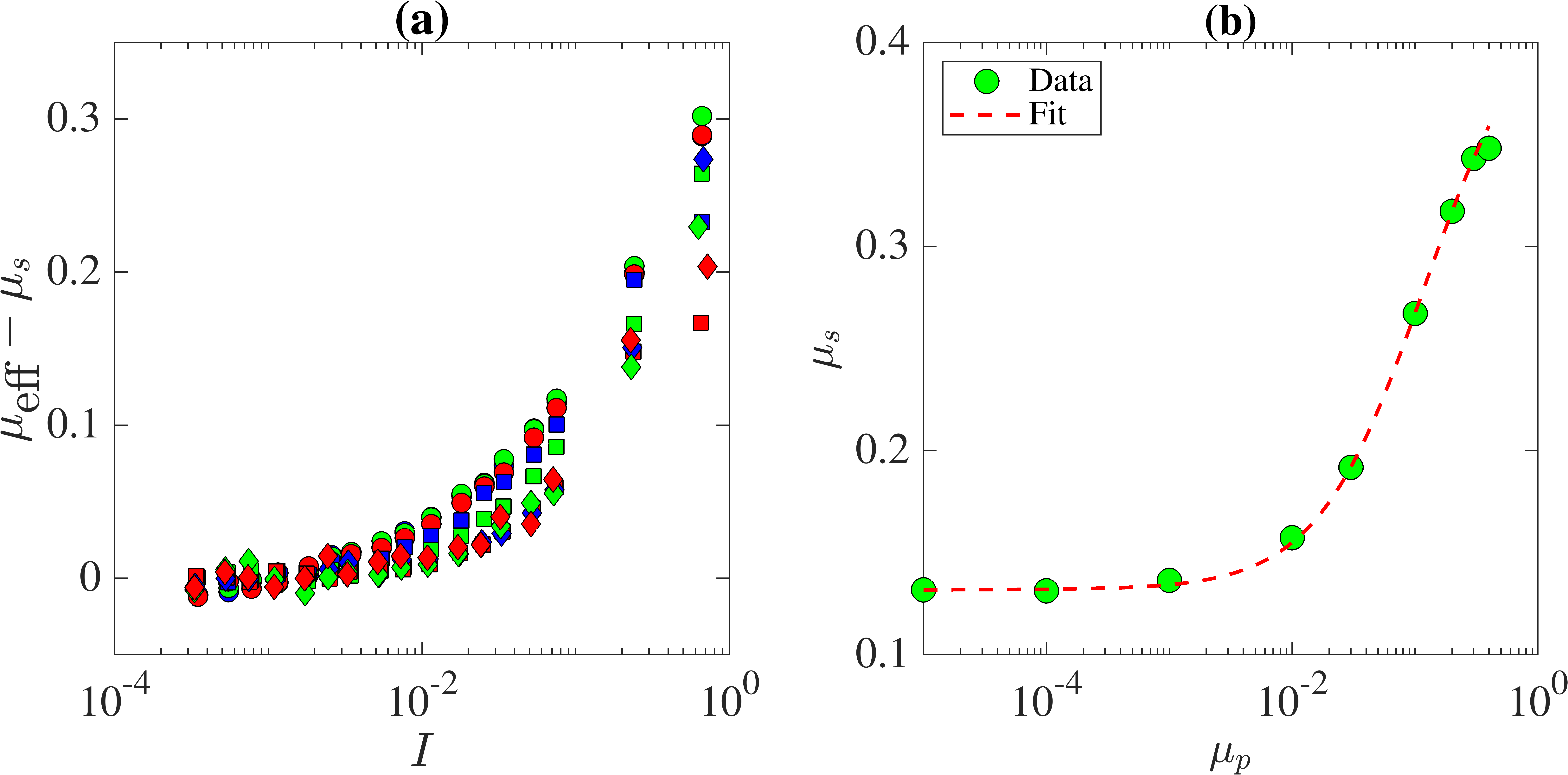}
  \caption{Influence of inter-particle friction on the effective frictional coefficient: (a) Relationship between $\mu_{\textrm{eff}} - \mu_s$ and $I$. $\mu_s$ is dependent on the inter-particle frictional coefficient. The markers are the same as those in Fig. \ref{fig2}(b); (b) We plot the relationship between $\mu_s$ and $\mu_p$ and fit the relationship with an equations denoted by red dashed curves (Eq.\ref{eq-mus1}).}
  \label{fig3}
\end{figure}

Based on the data in Fig. \ref{fig2}(a), for each set of simulations with the same $\mu_p$, we can fit an equation with the form of Eq. \ref{eq-muI}. Then, each set of simulations has one fitted static frictional coefficient, $\mu_s$. We subtract $\mu_s$ from $\mu_{\textrm{eff}}$, and plotted $\mu_{\textrm{eff}}-\mu_s$ against the inertial number $I$ in Fig. \ref{fig3}(a). On one hand, we do not expect a good collapse of the data in Fig. \ref{fig3}(a) since we can observe, in Fig. \ref{fig2}, that changing inter-particle friction also influences the transitional inertial number. In Fig. \ref{fig3}(a), as we increase the inertial number, $\mu_{\textrm{eff}}$ of systems with a smaller $\mu_p$ starts to increase earlier than that of systems with a larger $\mu_p$. On the other hand, we can obtain a relationship between $\mu_s$ and $\mu_p$ according to our simulation results. We have stated that $\mu_s$, although being fitted and acting as the minimum effective frictional coefficient when $I\rightarrow 0$, can be regarded as the static effective frictional coefficient of the system. 

In this study, $\mu_s$ only depends on $\mu_p$, thus, we plot the relationship between $\mu_s$ and $\mu_p$ in Fig. \ref{fig3}(b). When $\mu_p \leq 10^{-3}$, $\mu_s$ does not change much with respect to the increase of $\mu_p$. However, once $\mu_p$ is larger than $10^{-3}$, increasing $\mu_p$ results in a quick increase of $\mu_s$. This result is consistent with previous works that, in Ref. \cite{midi2004}, increasing $\mu_p$ levels up the $\mu_{\textrm{eff}}(I)$ relation, while $\mu_p = 0$ corresponds to $\mu_s\approx 0.1$. Hatano \cite{Hatano2007,Hatano2010} also found that increasing particle frictional coefficient from 0 to 1.0 results in an increase of $\mu_s$ from $\approx 0.1$ to $\approx 0.36$, and meanwhile, $\mu_s$ does not change much when $\mu_p\geq 0.3$. With our simulation results in Fig. \ref{fig3}(b), we can fit the $\mu_s(\mu_p)$ relationship with the following equation [read dashed curve in Fig. \ref{fig3}(b)],
\begin{linenomath*}
\begin{equation}
    \begin{split}
        \mu_s = \mu_c + \Delta\mu/\left( 1 + \mu_0/\mu_p\right), \label{eq-mus1}
    \end{split}
\end{equation}
\end{linenomath*}
where $\mu_c\approx 0.13$ is the static effective frictional coefficient of frictionless granular systems, which is only due to normal collisions between contacting particle pairs, $\Delta\mu\approx 0.28$ and $\mu_0 \approx 0.1$ are fitting parameters. The R-square of Eq.\ref{eq-mus1} is $\approx 0.998$. For, Eq.\ref{eq-mus1}, one thing that concerns us is that $\mu_c$, in this study, is approximately equal to 0.13, which is obviously larger than 0.1 obtained by previous research \cite{midi2004,Hatano2010,Azema2015internal}. This might be because that the inertial number $I$ of simulations with $\mu_p = 10^{-5}$, $10^{-4}$, and $10^{-3}$ is not small enough to reach the flat region of the $\mu_{\textrm{eff}}(I)$ relationship. It is clear that, for cases with $\mu_p = 10^{-5}$, $10^{-4}$, and $10^{-3}$, as we keep decreasing $I$, the effective frictional coefficient is continuously decreasing, even when $I$ reaches $10^{-4}$, $\mu_{\textrm{eff}}$ still has the tendency to decrease. This situation may result in an inaccurate prediction of $\mu_s$ for these sets of simulations. Nevertheless, our results quantitatively relate the static effective frictional coefficient and the microscopic friction between particles. Equation \ref{eq-mus1} also implies that, when $\mu_p\rightarrow +\infty$, the maximum $\mu_s$ a dry granular system can get is approximately 0.41, which is also consistent with previous works \cite{Hatano2007,Hatano2010}.

\begin{figure}[ht]
  \includegraphics[scale = 0.35]{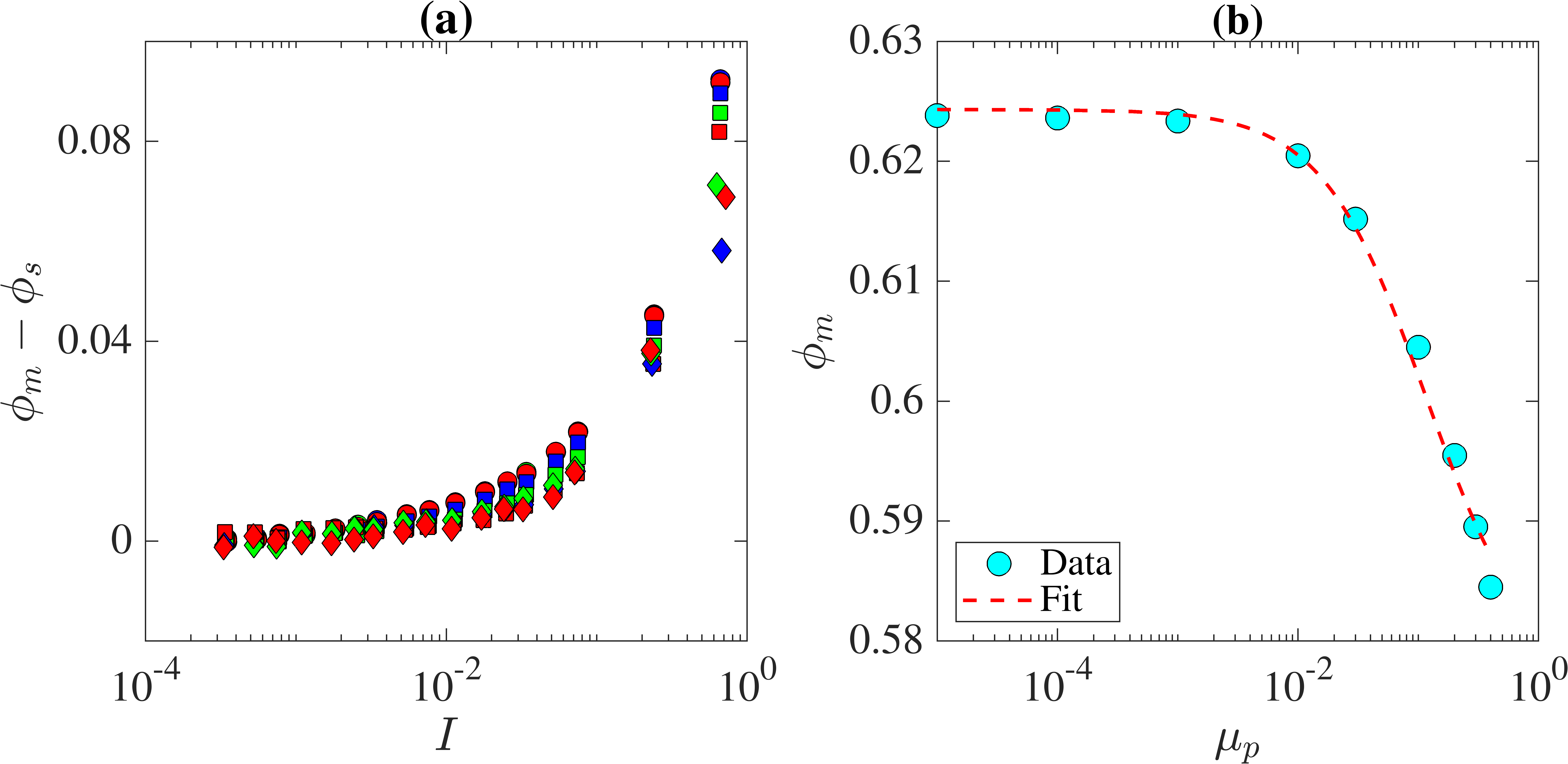}
  \caption{Influence of the inter-particle frictional coefficient on the solid fraction of the system: (a) Relationship between $\phi_m - \phi_s$ and $I$, where $\mu_s$ is dependent on the inter-particle frictional coefficient. The markers are the same as those in Fig. \ref{fig2}(b); In Fig. (b), we plot the relationship between $\phi_m$ and $\mu_p$ and fit the relationship with Eq.\ref{eq-phim1} (red dashed curve).}
  \label{fig4}
\end{figure}

Another aspect of the $\mu(I)$ rheology is that the solid fraction of a granular system can also be related to the inertial number, $I$. For frictionless granular systems, the solid fraction of a sheared granular assembly follows $\phi_s = \phi_J/(1+\beta I^{\alpha})$, where $\phi_J$ is the jamming solid fraction of frictionless systems, while, for frictional granular systems, we have $\phi_s = \phi_m/(1+\beta I^{\alpha})$, where $\phi_m<\phi_J$ and $\phi_m$ is the maximum solid fraction a sheared granular system can reach with a given inter-particle frictional coefficient. In Fig. \ref{fig4}(a), we plot $\phi_m - \phi_s$ against $I$ as if we flip and translate each curve so that its minimum is approximately 0. Similar to the relationship between $\mu_{\textrm{eff}} - \mu_s$ and $I$, data in Fig. \ref{fig4}(a) do not collapse onto on curve, which implies that subtracting $\phi_m$ from each curve does not eliminate the influence of the inter-particle friction.

Thus, we explore the relationship between $\phi_m$ and $\mu_p$, and plot their relationship in Fig. \ref{fig4}(b). As we increase the inter-particle frictional coefficient, the maximum solid fraction $\phi_m$ decreases, and $\phi_m$ is within the range of the random loose packing fraction \cite{song2008phase}, $\phi_{RLP}\approx 0.54$, and the random close packing fraction of frictionless systems \cite{song2008phase}, $\phi_{RCP}\approx 0.636$. We fit the $\phi_m(\mu_p)$ relationship with the following equation,
\begin{linenomath*}
\begin{equation}
    \begin{split}
        \phi_m = \phi_1 - {\Delta\phi}/(1 + \mu_0/\mu_p),\label{eq-phim1}
    \end{split}
\end{equation}
\end{linenomath*}
where $\phi_1\approx 0.625$, $\Delta\phi\approx 0.048$, $\mu_0\approx 0.1$ is the same as that in Eq.\ref{eq-mus1}. Equation \ref{eq-phim1} can clearly provide us a good fit, but we have to note that, similar to $\mu_s$, the $\phi_m$'s we obtained from cases with $\mu_p = 10^{-5}, 10^{-4}$, and $10^{-3}$ might not be accurate since the inertial number is not small enough.

\subsection{Dimensional analysis and frictional number}

\begin{figure}[ht]
  \includegraphics[scale = 0.35]{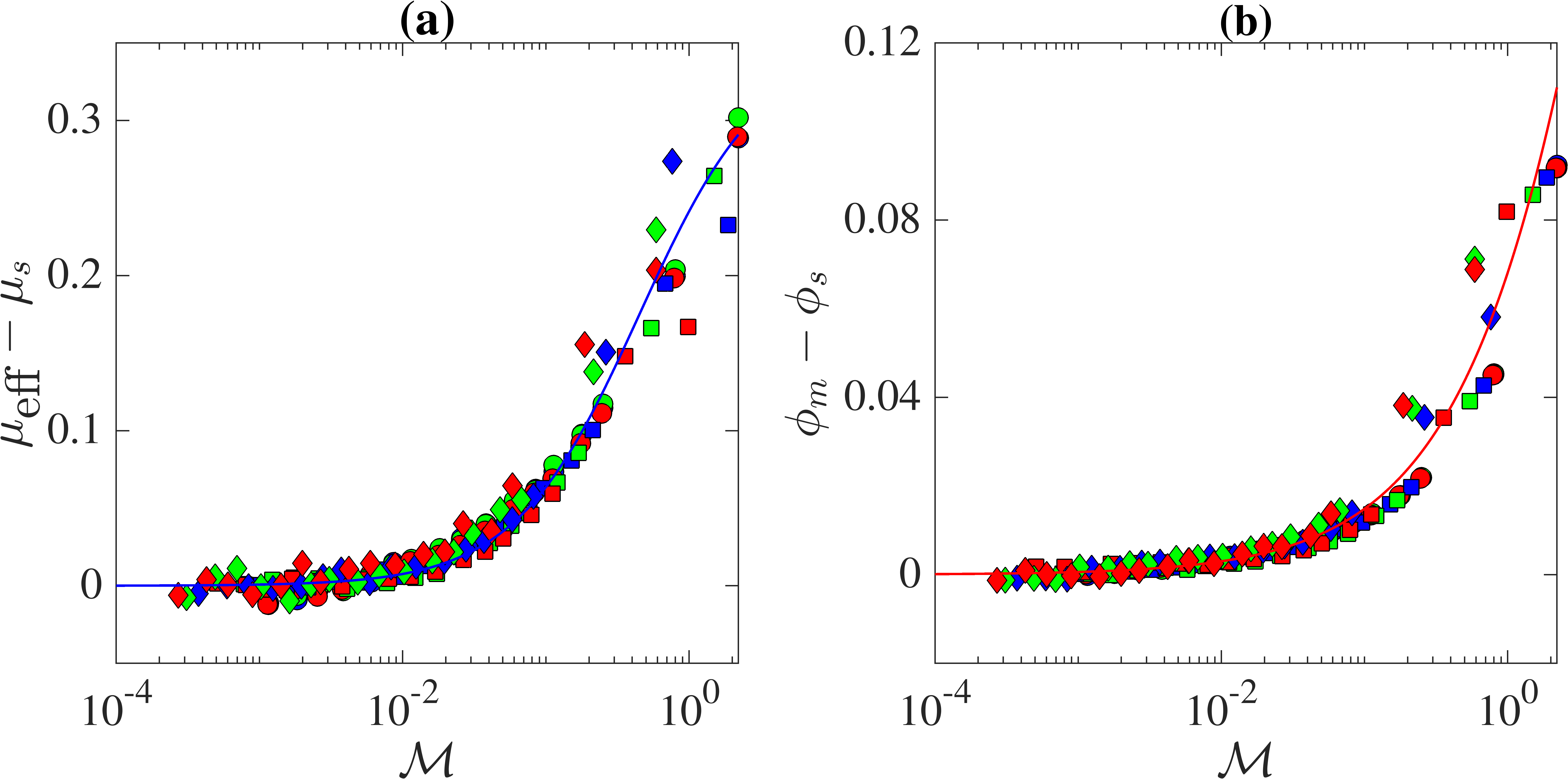}
  \caption{Rheological relationship with new dimensionless number, $\mathcal{M}$: (a) Relationship between $\mu_{\textrm{eff}} - \mu_s$ and $\mathcal{M} = I/\sqrt{\mu_{\textrm{fl}} + \beta\mu_p}$; (b) Relationship between $\phi_sm - \phi_s$ and $\mathcal{M}$. The solid curves in both Figs. (a) and (b) represent fitted equations, respectively. The markers are the same as those in Fig. \ref{fig2}(b)}
  \label{fig5}
\end{figure}

We can see that, in Fig. \ref{fig3}(a) and \ref{fig4}(a), tuning the $y-$axis has difficulties in collapsing the data. The inter-particle friction can not only influence both the effective aspect ratio and the solid fraction but also influence the transition from quasi-static flow regime to inertial flow regime. In previous research, the inertial number, $I$, is usually interpreted as the ratio between the relaxation time scale, $d/\sqrt{\sigma_n/\rho}$, and the shearing (macroscopic) time scale, $1/\dot{\gamma}$, which conveniently neglect the influence of the inter-particle friction. However, from a viewpoint of stresses, we tend to interpret the inertial number as a ratio between two different theoretical stresses, i.e. driving stress (which scales with the theoretical inertial stress, $\sigma_i$) and resistant stress (which scales with the theoretical frictional stress, $\sigma_r$), thus,
\begin{linenomath*}
\begin{equation}
\begin{split}
    \sigma_i=\rho\dot{\gamma}^{2}d^2,\ \sigma_r=\mu_{g}\sigma_n,
\end{split}
\end{equation}
\end{linenomath*}
where $\mu_g$ is a general frictional coefficient, which is predicted to be different from $\mu_p$. Different from the classic inertial number, $I$, the ratio between $\sigma_i$ and $\sigma_r$ incorporates the influence of friction, thus, we can write the new dimensionless number as,
\begin{linenomath*}
\begin{equation} \label{eq-IM}
    \begin{split}
        \mathcal{M} = \sqrt{\sigma_i/\sigma_r} = \sqrt{\rho\dot{\gamma}^{2}d^2/(\mu_{g}\sigma_n)} = I/\sqrt{\mu_g},
    \end{split}
\end{equation}
\end{linenomath*}
and name it as the frictional number. The general frictional effect should include two aspects: (i) the particle inter-lock effect due to frictionless interactions, and (ii) the frictional effect because of the existence of $\mu_p$. Thus, we hypothesize that $\mu_g$ can be written as a linear combination of effective friction of frictionless granular systems and the inter-particle frictional coefficient so that,
\begin{linenomath*}
\begin{equation}
    \begin{split}
        \mu_g = \mu_{\textrm{fl}} + \beta\mu_p,
    \end{split}
\end{equation}
\end{linenomath*}
where $\mu_{\textrm{fl}} \approx 0.1$ is the effective frictional coefficient when $\mu_p = 0$ and $I = 0$, which is also consistent with previous works \cite{Hatano2007,Hatano2010,Azema2015internal}. The linear combination of $\mu_{\textrm{fl}}$ and $\mu_p$ is similar to our previous work of granular column collapses, where we define the effective aspect ratio using the combination of boundary friction and inter-particle frictions\cite{man2021deposition,man2020finite}. To acquire a better collapse of all the data, we take $\beta \approx 3.5$. In Fig.\ref{fig5}(a), we plot the relationship between $\mu_{\textrm{eff}} - \mu_s$ and $\mathcal{M} = I/\sqrt{\mu_{\textrm{fl}} + \beta\mu_p}$. In Fig.\ref{fig5}(b), we plot the relationship between $\phi_m - \phi_s$ and $I_M$. After using $\mathcal{M}$ as the $x-$axis, both relationships are clearly improved, and we can obtain following rheological equations similar to the $\mu(I)$ relationship,
\begin{linenomath*}
\begin{subequations}
\begin{align}
    \mu_{\textrm{eff}} - \mu_s = \Delta\mu_{m}/\left( 1 + \mathcal{M}_0/\mathcal{M} \right), \label{eq-muIM}\\
    \phi_m - \phi_s = \phi_J - \phi_J/\left( 1 + \beta_{\phi} \mathcal{M}^{\alpha} \right),
\end{align}
\end{subequations}
\end{linenomath*}
where $\phi_J = 0.634$ is a classical jamming solid fraction of frictionless systems reported by Ref. \cite{song2008phase}, $\Delta\mu_{m}\approx 0.35$, $\mathcal{M}_0 \approx 0.45$, $\beta_{\phi}\approx 0.12$, and $\alpha\approx 0.7$. Thus, both the effective frictional coefficient, $\mu_{\textrm{eff}}$, and the solid fraction, $\phi_s$, can be completely described by the frictional number, $\mathcal{M}$, and the inter-particle frictional coefficient, $\mu_p$. In both Figs. \ref{fig5}(a) and \ref{fig5}(b), when $\mathcal{M} \gtrapprox 0.2$, the data points become rather scattered. These data points correspond to cases with large inertial effects, which indicates that the system is approaching a granular gas regime where fluctuations is pervasive.

\begin{figure}[ht]
\centering
  \includegraphics[scale = 0.35]{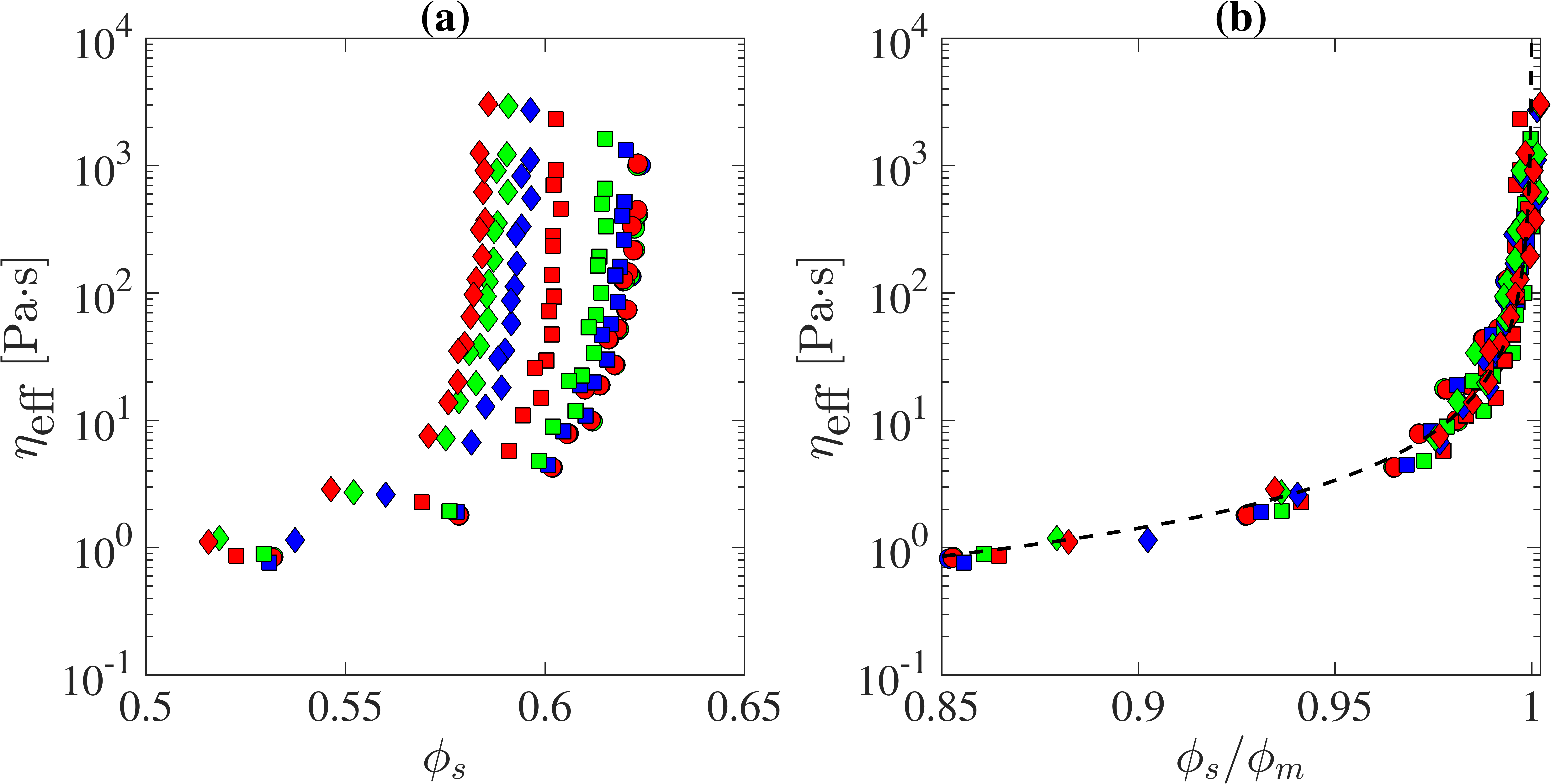}
  \caption{Effective viscosity of granular systems: (a) Relationship between the effective viscosity, $\eta_{\textrm{eff}}$, and solid fraction, $\phi_s$; (b) Relationship between $\eta_{\textrm{eff}}$ and $\phi_s/\phi_m$, where $\phi_m$ depends on $\mu_p$. The dashed curve in Fig. (b) represents the fitted equation. The markers are the same as those in Fig. \ref{fig2}(b)}
  \label{fig6}
\end{figure}

\begin{figure}
    \centering
    \includegraphics[scale = 0.35]{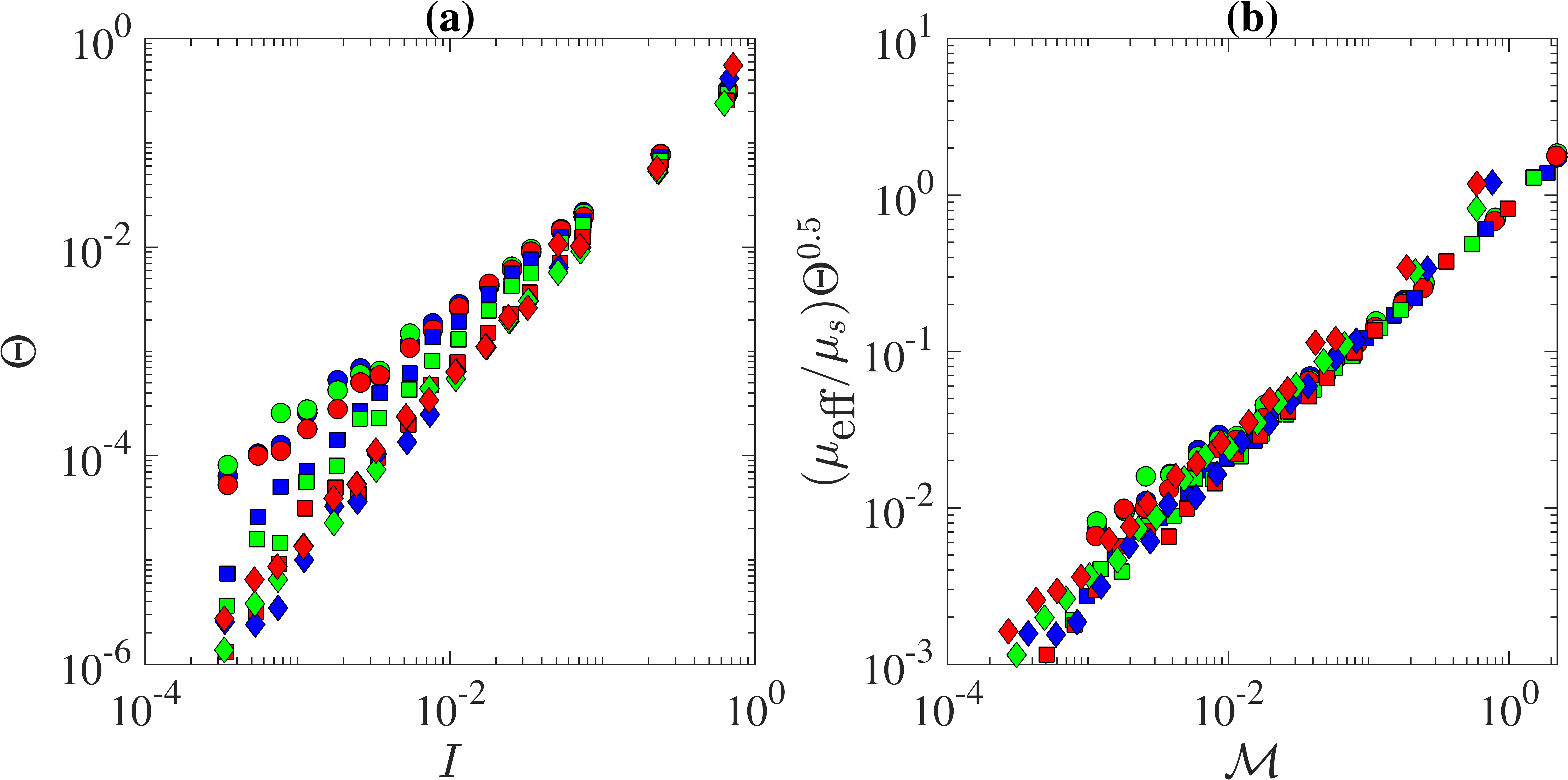}% Here is how to import EPS art
    \caption{Dimensionless granular temperature of granular systems: (a) The relationship between $\Theta = \rho T_g/\sigma_n$ and $I$; (c) Relationship between $(\mu_{\textrm{eff}}/\mu_s)\Theta^{0.5}$ and $\mathcal{M}$. The markers are the same as those in Fig. \ref{fig2}(b)}
    \label{fig7}
\end{figure}

Additionally, in order to simulate the behavior of granular flows as a complex fluid, we investigate the effective viscosity, $\eta_{\textrm{eff}} = \tau/\dot{\gamma}$, and its relationship with the solid fraction. Fig. \ref{fig6}(a) plots $\eta_{\textrm{eff}}$ against the solid fraction, and obviously, we do not obtain a good collapse of data since $\eta_{\textrm{eff}}$ approaches infinity at different solid fractions for cases with different $\mu_p$. However, for simulation results with the same $\mu_p$, $\eta_{\textrm{eff}}$ scales exponentially with respect to $\phi_s$. Since $\phi_m$ can be represented by Eq. \ref{eq-phim1}, we eliminate the influence of $\phi_m$ by changing the $x-$axis from $\phi_s$ to $\phi_s/\phi_m$, with which, all the data collapse seemingly onto one curve that can be fitted with the following equation,
\begin{linenomath*}
\begin{equation} \label{eq-visco}
    \begin{split}
        \eta_{\textrm{eff}} = 0.08\left( 1 + \phi_s/\phi_m \right)^{-\zeta}
    \end{split}
\end{equation}
\end{linenomath*}
where $\zeta = 1.25$ is a fitted parameter. The form of Eq. \ref{eq-visco} is similar to the relationship between $\eta_{\textrm{eff}}$ and $\phi_s/\phi_m$ for granular suspensions reported in Ref. \cite{guazzelli2018}. However, for granular suspensions, the scaling exponent $\xi$ is often equal to 2, which shows that the existence of interstitial fluid enhances the inter-particle contact, and helps to achieve larger increase of the effective viscosity with respect to the increase of the solid fraction.

\subsection{Granular temperature and regime transition}

In previous sections, we establish a friction-dependent constitutive framework for frictional granular systems using the frictional number, $\mathcal{M}$, and inter-particle frictonal coefficient, $\mu_p$. However, this constitutive framework depends largely on fitting, especially the fitted equation of both $\mu_s(\mu_p)$ relationship and $\phi_m(\mu_p)$ relationship. Inspired by Kim et al. \cite{Kim2020power}, we seek to utilize the concept of the granular temperature of sheared granular systems. We define the granular temperature of as $T_g\equiv\langle u_i^{\prime}u_i^{\prime} \rangle/3$, where $u_i^{\prime}=u_i - \langle u_i \rangle$ is the $i$th component of the velocity fluctuation, and $\langle\cdot\rangle$ denotes the calculation of averages. Same as Kim et al. \cite{Kim2020power}, the dimensionless granular temperature is $\Theta = \rho T_g/\sigma_n$. If we plot $\Theta$ against $I$, as shown in Fig. \ref{fig7}(a), $\Theta$ scales with $I^{\xi}$ for simulations with same $\mu_p$. However, the scaling parameter, $\xi$, depends on the inter-particle frictional coefficient, $\mu_p$. When $\mu_p = 10^{-5}$, $\xi\approx 1$, but when $\mu_p = 0.4$, $\xi$ becomes approximately 2. The dimensionless granular temperature is determined not only by the inertial number but also by the inter-particle friction. For a given inertial number, $I$, larger $\mu_p$ corresponds to a smaller $\Theta$. All the data gradually converge into one power-law relationship when $\Theta \gtrapprox 10^{-2}$.

\begin{figure}[ht]
  \includegraphics[scale = 0.35]{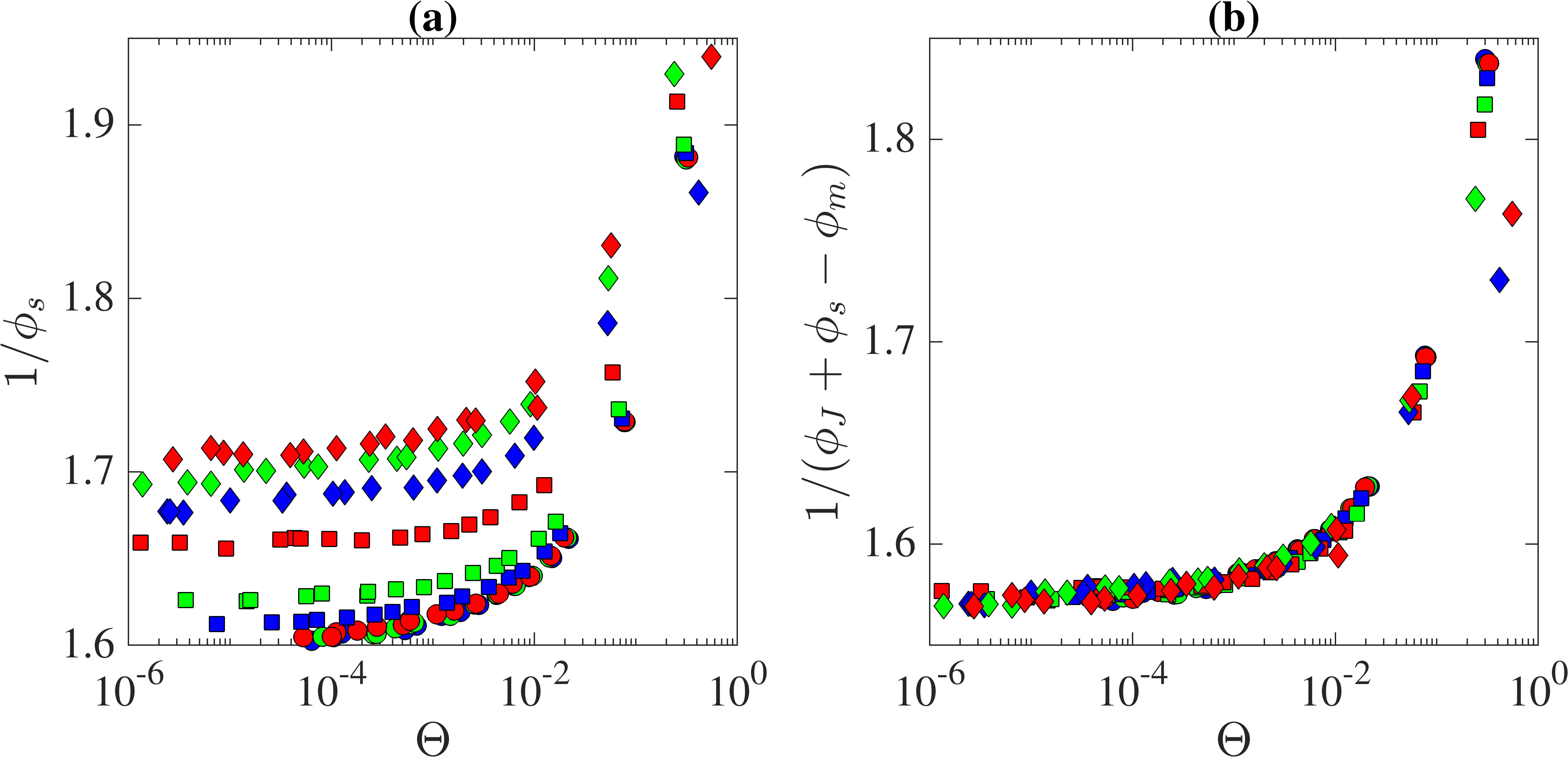}
  \caption{Inversed solid fraction of granular systems: (a) Relationship between the dimensionless granular temperature, $\theta$, and the inversed solid fraction, $1/\phi_s$, of the system; (b) The relationship between $\theta$ and $1/(\phi_J+\phi_s-\phi_m)$. The markers are the same as those in Fig. \ref{fig2}(b)}
  \label{fig8}
\end{figure}

According to Fig. \ref{fig7}(a), the relationship between $\Theta$ and $I$ depends on the inter-particle friction. Similar to Kim et al. \cite{Kim2020power}, We then change the $x-$axis into the frictional number, $\mathcal{M}$, and change the $y-$axis to $(\mu_{\textrm{eff}}/\mu_s)\Theta^{0.5}$, and plot their relationship between in Fig. \ref{fig7}(b). Surprisingly, rescaling the data with $\mathcal{M}$ and $(\mu_{\textrm{eff}}/\mu_s)\Theta^{0.5}$ can reshape the $\Theta(I)$ relationship of systems with different $\mu_p$ into approximately a power-law scaling. This indicates that the dimensionless granular temperature, $\Theta$, reflects not only the inertial effect denoted by the inertial number, $I$, but also the frictional interaction between particles. 

We note that the exponential parameter of $\Theta$, in our study, is 0.5, which is different from Ref. \cite{Kim2020power}, where they stated that, for a 3D system, the exponential parameter was approximately $1/6$. In Ref. \cite{Kim2020power}, $\Theta$ was account for the influence of fluidity field and its corresponding fluctuation difference. However, in our study, differences in velocity fluctuations come from varying inter-particle friction, $\mu_p$, thus, the scaling parameter is naturally different from the one obtained in Ref. \cite{Kim2020power}. The power-law relationship between $(\mu_{\textrm{eff}}/\mu_s)\Theta^{0.5}$ and $\mathcal{M}$ helps us establish another constitutive framework that the effective frictional coefficient, $\mu_{\textrm{eff}}$, can be calculated with both the dimensionless granular temperature, $\Theta$, and the frictional number, $\mathcal{M}$, which incorporates the influence of the inter-particle frictional coefficient.

\begin{figure}[ht]
  \includegraphics[scale = 0.35]{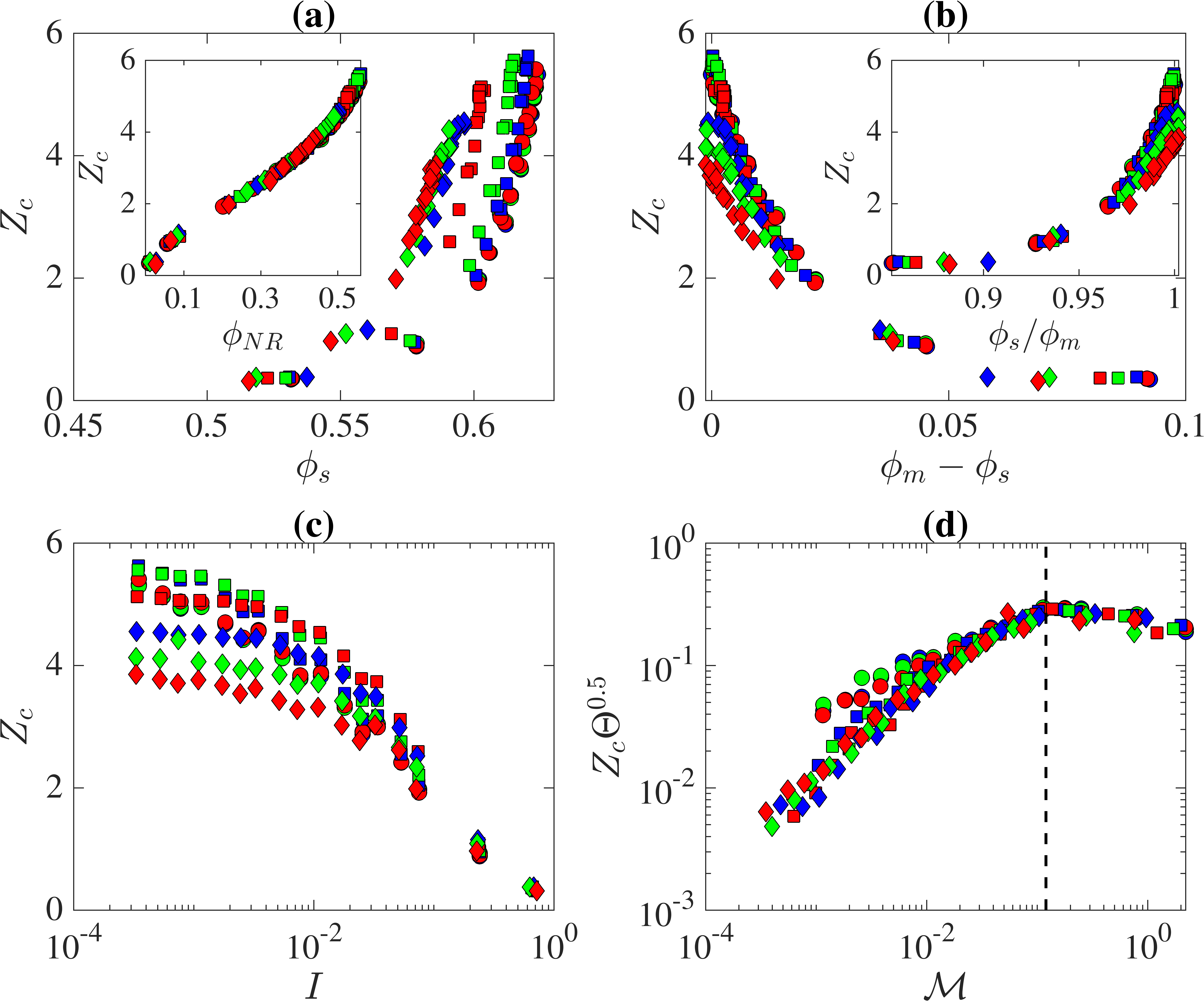}
  \caption{Average coordination number of granular systems: (a) Relationship between the average coordination number, $Z_c$, and the solid fraction, $\phi_s$, of the system; (b) the relationship between $Z_c$ and $\phi_m - \phi_s$; (c) the relationship between $Z_c$ and $I$; and (d) the relationship between $Z_c\Theta^{0.5}$ and $\mathcal{M}$. The markers are the same as those in Fig. \ref{fig2}(b). The inset of Fig. (a) plots the relationship between $Z_c$ and the non-rattler solid fraction, $\phi_{NR}$, and the inset of Fig. (b) shows the relationship between $Z_c$ and $\phi_s/\phi_m$}
  \label{fig9}
\end{figure}

\begin{figure*}
\centering
\includegraphics[scale = 0.5]{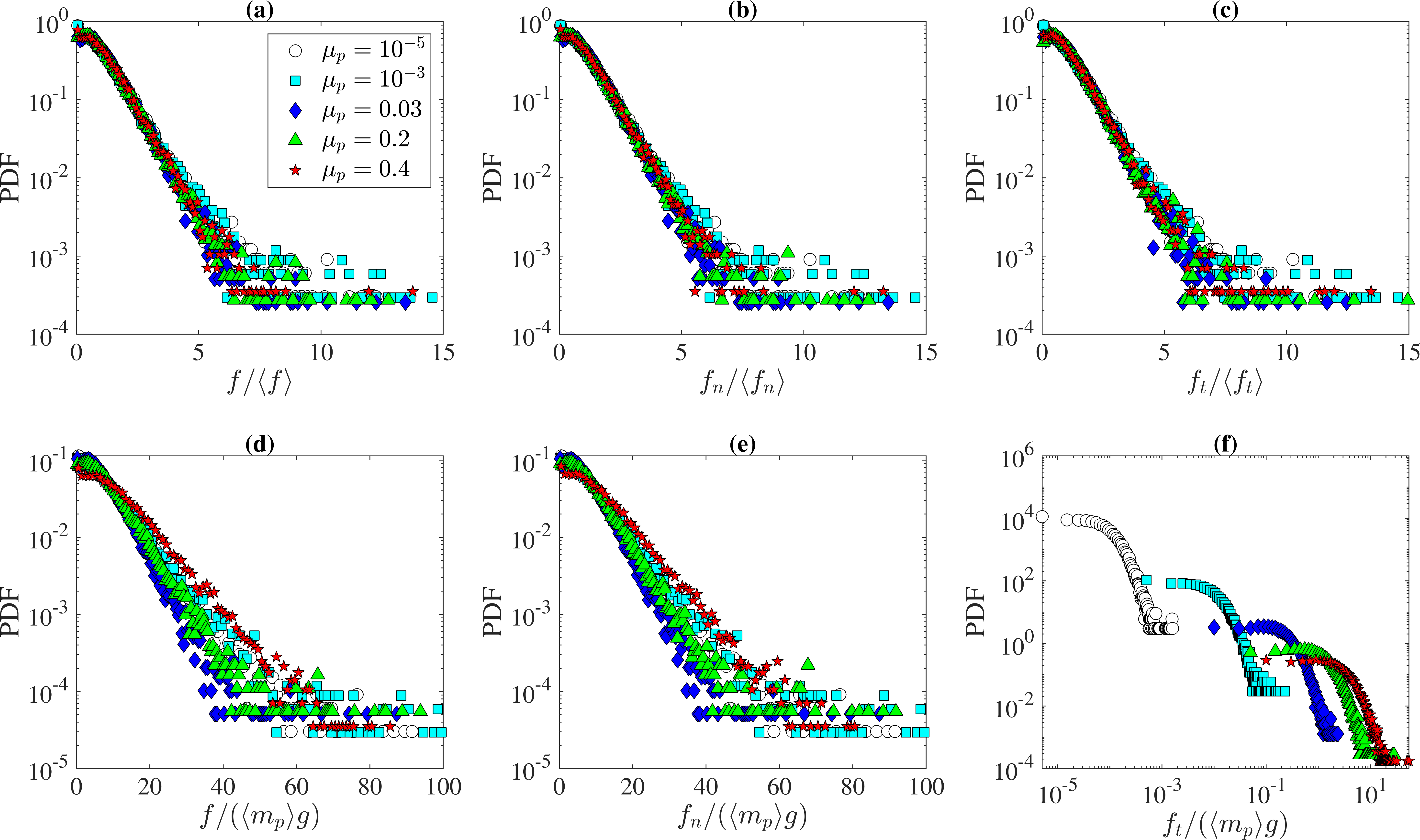}% Here is how to import EPS art
\caption{Histograms of contact forces: Histograms of (a) total contact forces, (b) normal contact forces, and (c) tangential contact forces, all normalized by their averages, when $\sigma_n = 100$ Pa, $\dot{\gamma} \approx 5.7\times 10^{-3}$; Histograms of (d) total contact forces, (e) normal contact forces, and (f) tangential contact forces, all normalized by the average particle weight, $\langle m_p\rangle g$.}
\label{fig10}
\end{figure*}

We then plot the relationship between the inverse solid fraction, $1/\phi_s$, and the dimensionless granular temperature, $\Theta$, in Fig. \ref{fig8}(a). Mari et al \cite{mari2009jamming,liu2010jamming} compared the granular jamming transition and glass transition and found that, similar to other amorphous systems, granular material also exhibits regimes that can be classified as liquid-like states and metastable glassy states, and exists a set of J-points that can be regarded as jamming transition points. In our analysis, changing $\mu_p$ can also provide us with a set of metastable states shown in Fig. \ref{fig8}(a), where $1/\phi_s$ is stable when $\Theta$ varies from $10^{-6}$ to $\approx 10^{-2}$. When $\Theta > 10^{-2}$, the system gradually transit to a liquid-like state similar to what is reported in Ref. \cite{mari2009jamming}. Rescaling the $y-$axis to $1/(\phi_J+\phi_s-\phi_m)$ (so that $\textrm{max}(\phi_J+\phi_s-\phi_m)$ for each set of simulations with the same $\mu_p$ is equal to $\phi_J$), we then plot $1/(\phi_J+\phi_s-\phi_m)$ against $\Theta$, which shows that the transition from a meta-stable state to a liquid-like state is relatively universal with respect to $\Theta$, and $\Theta\approx 10^{-2}$ marks this transition. The transitional $\Theta \approx 10^{-2}$, is also similar to the transitional point in Fig. \ref{fig7}(a) and Ref. \cite{Kim2020power}, where the $\Theta(I)$ relationship always converges to one curve when $\Theta\gtrapprox 10^{-2}$ regardless of loading conditions.

\subsection{Coordination number and contact statistics}

In Fig. \ref{fig9}, we investigate the scaling of the average coordination number and its dependence on the solid fraction and the inertial number of granular systems. We calculate the coordination number, $Z_c$, as the average contact number in a granular system. Fig. \ref{fig9}(a) and (b) show that, when $\phi_s$ is small, the coordination number, $Z_c$, converges to 0. This is when the system is subjected to low pressure or high shear rate that binary collisions dominate the dynamics of the assembly, and the system is in a liquid-like or even a gas-like state. However, when the solid fraction approaches $\phi_m$, which is the maximum solid fraction a sheared system can reach for certain $\mu_p$, $Z_c$ often converges to different values. As shown in Fig. \ref{fig9}(b) and its inset figure, $Z_c$ converges to $\approx 6$ when $\phi_s$ is approaching $\phi_m$ for simulations with $\mu_p = 10^{-5}$, but it converges to only $\approx 4$ for simulations with $\mu_p = 0.4$. Increasing inter-particle frictional coefficient greatly decreases the ability of granular systems to build concrete force networks in the system. From another viewpoint, a system can still reach a metastable state with much lower coordination numbers if the inter-particle frictional coefficient is sufficiently large. Fig. \ref{fig9}(c) shows us similar results but using the inertial number, $I$, as the critical parameter. Similarly, $Z_c$ plateaus at different levels for systems with different inter-particle frictions, when $I$ is decreasing.

Since the relationship between $\phi_s$ and $Z_c$ is scattered, we calculate the non-rattler solid fraction, $\phi_{NR}$, of granular systems to test if there exists a collapsed relationship between $Z_c$ and $\phi_{NR}$, and plot the results in the inset of Fig. \ref{fig9}(a). We define $\phi_{NR}$ as the solid fraction of particles with no less than 3 contacts with their neighbors. As shown in the inset of Fig. \ref{fig9}(a), even though friction influences the average coordination number dramatically, it does not break the universality in terms of the $Z_c(\phi_{NR})$ relationship. This implies that changing the frictional coefficient influences the number of non-rattler particles and their corresponding contact structure. 

Considering the influence of inter-particle friction and its corresponding velocity fluctuation, we introduce the frictional number, $\mathcal{M}$, and the dimensionless granular temperature, $\Theta$, into the analysis of $Z_c$, and plot the relationship between $Z_c\Theta^{0.5}$ and $\mathcal{M}$ in Fig. \ref{fig9}(d). We not only obtain a well-behaved collapse of all the data but also find another transitional point presented in the $Z_c\Theta^{0.5}-\mathcal{M}$ relationship. This transition occurs at $\mathcal{M}\approx 10^{-1}$, which is close to the transition point where $\mu_{\textrm{eff}}$ starts to increase and becomes more scattered. The physics of this transition needs further investigations in future studies.

Additionally, we look into the statistics of contact forces in sheared granular systems with different inter-particle frictional coefficients, and plot histograms of normalized contact forces, $f/\langle f\rangle$, normalized normal contact forces, $f_n/\langle f_n\rangle$, and normalized tangential contact forces, $f_t/\langle f_t\rangle$, in Fig. \ref{fig10}(a - c). These granular systems are subjected to the same loading condition, where $\sigma_n = 100$ Pa and $\dot{\gamma} \approx 5.7\times 10^{-3}$. The only difference lies in the interparticle frictional coefficient. Different from what we have observed in the previous sections and figures that $\mu_p$ subject great influence to the granular system, the histograms of $f/\langle f\rangle$, $f_n/\langle f_n\rangle$, and $f_t/\langle f_t\rangle$ do not differ much from each other as we change their inter-particle frictional coefficients. All these histograms exhibit a similar exponential decay. However, a subtle difference can be observed that simulations with a smaller $\mu_p$ often have higher chances to find contact forces larger than 5 times the average.

We also plot the histograms of total contact forces, normal contact forces, and tangential contact forces without being normalized by their averages (but normalized by the average particle weight, $\langle m_p\rangle g$) in Fig. \ref{fig10}(d - f). As we increase $\mu_p$ from $10^{-5}$ to 0.4, it is natural that the histogram of $f_t$ moves from left to right. The histograms scale exponentially with respect to the However, the histograms of $f$ and $f_n$ behave different from those of $f/\langle f\rangle$ and $f_n/\langle f_n\rangle$, even though their loading condition is the same. Increasing $\mu_p$ from $10^{-5}$ to 0.4, the slope of the histograms of $f$ and $f_n$ first increase then decrease. This shows that systems tend to have larger contact forces when particles are either almost frictionless or very frictional. We can also link back this behavior to the phase diagram suggested by DeGiuli et al.\cite{degiuli2016} shown in Fig. \ref{fig1}. When the inertial number is fixed, increasing interparticle frictional coefficient

\section{Conclusions}
\label{sec-conclu}
In this study, to clarify the influence of inter-particle frictional coefficient, $\mu_p$, on the rheology of granular materials, we vary $\mu_p$ from $10^{-5}$ to 0.4. The results show that changing inter-particle friction can not only influence $\mu_{\textrm{eff}}$ by shifting the $\mu(I)$ rheology curve upward, but also change the transitional inertial number, indicating that increasing $\mu_p$ leads to an increase of both $\mu_{\textrm{eff}}$ and $I_0$ of Equation \ref{eq-muI} and a decrease of $\phi_s$. We first investigate the influence of both $\mu_p$ on the static effective frictional coefficient, $\mu_s$, and maximum solid fraction a sheared system can reach, $\phi_m$, and summarize its influence as Eqs. \ref{eq-mus1} and \ref{eq-phim1}. We further introduce a new dimensionless number, $\mathcal{M}$, which can incorporate both the effect of inertial number, $I$, and inter-particle frictional coefficients, $\mu_p$. With the introduction of $\mathcal{M}$, we can establish a well-behaved rheological representation of dry granular systems with different inter-particle frictional coefficients. Nevertheless, the frictional number, $\mathcal{M}$, resembles the effective aspect ratio we have obtained to quantify the run-out behavior of granular column collapses in our previous work\cite{man2021deposition,man2020finite}, where we derived the effective aspect ratio as a ratio between inertial effect and frictional effect. The similarity between these two studies may indicate the universality of inertial/friction ratio in studies of granular systems.

Learning from Ref. \cite{Kim2020power}, we also include the dimensionless granular temperature, $\Theta$, in our analysis to consider the influence of $\mu_p$ in terms of velocity fluctuations. Even though the scaling of $\Theta$ with respect to $I$ varies given different $\mu_p$, the relationship between $(\mu_{\textrm{eff}}/\mu_s)\Theta^{0.5}$ and $\mathcal{M}$ collapses onto one power-law curve, which can provide us with a simpler friction-dependent constitutive relationship of sheared granular assemblies. Further analyses on the relationship between inverse solid fraction and dimensionless granular temperature confirm that $\Theta$ provides us with a relatively universal transition from a so-called metastable state to a liquid-like state at $\Theta\approx 10^{-2}$. We can also obtain the scaling of the coordination number, $Z_c$, with respect to the frictional number, $\mathcal{M}$, with the assistance of $\Theta$, which implies that the dimensionless granular temperature, $\Theta$, can help quantify the influence of both frictions and fluctuations. However, we note that, in this paper, the choices of $\mu_p$ are limited within $10^{-5}$ and $0.4$. To obtain a more thorough investigation, we have to further decrease and increase the inter-particle frictional coefficient in the future. Besides, this work only investigates the behavior of sheared granular assemblies of spherical particles with narrowly and uniformly distributed particle sizes. This choice of particle shape and size distribution limits the number of possible microscopic topological structures one can observe in a granular system, which should be studied and analyzed in future works.

\acknowledgement{The authors acknowledge the financial support from the General Program (NO. 12172305) of the National Natural Science Foundation of China and Westlake University, and thank the Westlake High-Performance Computing Center for computational resources and related assistance. The authors would like to thank Prof. Ling Li from Westlake University and Prof. Herbert Huppert from the University of Cambridge for helpful discussions related to this study.}

%%%%%%%%%%%%%%%%%%%%%%%%%%%%%%%%%%%%%
%Reference
\bibliographystyle{spmpsci}
\bibliography{rheologyPaperBib.bib}

%\end{sloppypar}
\end{document}